\documentclass[twocolumn,aps]{revtex4-2}
\usepackage{amsmath}
\usepackage{amssymb}
\usepackage{lipsum}
\usepackage{graphicx}
\usepackage{dcolumn}
\usepackage{bm}
\usepackage{framed}
\usepackage{booktabs}
\usepackage{xcolor}
\usepackage{siunitx}
\usepackage{epstopdf}

\begin{document}

\preprint{Phys.Rev.B}

\title{ Magnetotransport in a 2D Hybrid Band System: Dirac and Heavy Hole Interplay}

\author{G. M. Gusev,$^1$  A. D. Levin,$^1$  V. A. Chitta,$^1$
 Z. D. Kvon,$^{2,3}$ and  N. N. Mikhailov$^{2,3}$}

\affiliation{$^1$Instituto de F\'{\i}sica da Universidade de S\~ao
Paulo, 135960-170, S\~ao Paulo, SP, Brazil}
\affiliation{$^2$Institute of Semiconductor Physics, Novosibirsk
630090, Russia}
\affiliation{$^3$Novosibirsk State University, Novosibirsk 630090,
Russia}

\date{\today}
\begin{abstract}
We investigate magnetoresistivity and the Hall effect in a 6.3 nm gapless HgTe quantum well - a two-dimensional hybrid band system featuring coexisting linear (Dirac-like) and parabolic hole energy bands at low energies. Using a classical two-subband model that includes intervalley scattering, we reveal a striking tenfold enhancement of the Hall resistance, mainly driven by the dominant transport contribution of Dirac holes. A comprehensive magnetotransport analysis allows us to extract key parameters, such as the mobilities of both carrier types, providing insight into their complex interplay. These results establish the HgTe quantum well as a distinctive platform for investigating novel transport phenomena in hybrid band systems, enhancing our understanding of mixed carrier magnetotransport.
\end{abstract}
\maketitle

\section{Introduction}
The recent discovery of the large magnetoresistance (MR) exceeding $100\%$ in conductors and semimetals under only a few tesla has garnered considerable attention \cite{niu}. This finding necessitates a reevaluation of the semiclassical theory, particularly in the context of the Boltzmann transport equation. The conventional Drude magnetoresistivity in metals is  exactly zero due to the compensation of the Lorentz force by the electric field resulting from the Hall effect.
\begin{figure*}
\includegraphics[width=16cm]{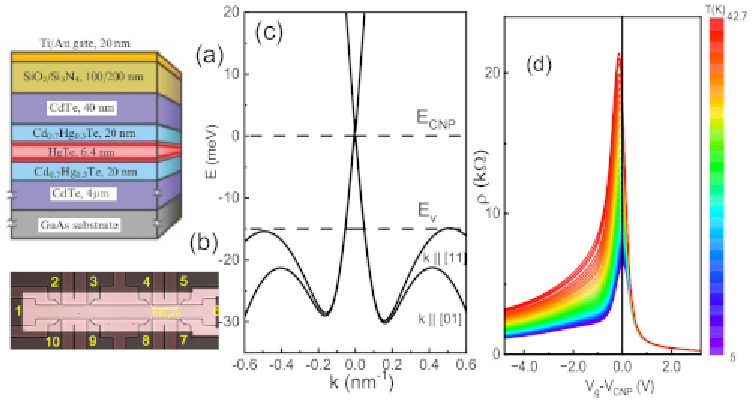}
\caption{ (a) Schematic of the transistor. (b) A top view of the sample. (c) Schematic representation of the energy spectrum of a 6.3-nm mercury telluride quantum well. (d) The resistivity of HgTe  quantum well as a function of the gate voltage for different temperatures (sample A).}
\label{sample}
\end{figure*}
The scenario undergoes a significant transformation when dealing with a system featuring the occupation of two or more subbands. In such cases, the anticipated magnetoresistance assumes a classical parabolic form, contingent upon the unique properties inherent to the system. In simpler situation, the magnetoresistance is governed by difference in subband mobilities within independent channels, as outlined in \cite{ziman}. In more practical situations, the inclusion of intersubband scattering becomes crucial, leading to a more intricate manifestation of magnetoresistance, as explained in detail in the model \cite{zaremba}. The experiments conducted in a GaAs double quantum well, representing a standard two-subband system, exhibit remarkable agreement with classical and quantum theory in describing low-field magnetoresistance \cite{fletcher, mamani, raichev}.

The magnetoresistance of a two-component system exhibits significant enhancement when the presence of both electrons and holes coexists, as opposed to a system characterized solely by size-quantized subbands. This phenomenon has been extensively investigated in two-dimensional semimetals, particularly in HgTe-based quantum wells \cite{kvon}. In such systems, the distinct mobility of electrons and holes results in a remarkable giant magnetoresistance even in relatively low magnetic fields. A notable illustration of this compensated magnetoresistance is evident in the observation of nonsaturating magnetoresistance in three-dimensional semimetals such as $WTe_{2}$  \cite{ali}, Dirac semimetal $Cd_3As_2$ \cite{liang}, and in Weyl semimetal NbP \cite{shekhar}.

The creation of thin films of $WTe_{2}$ with adjustable concentrations \cite{wang} has provided valuable insights. It has been demonstrated that the underlying mechanism driving this phenomenon is the compensation of electron-hole charges. At the charge neutrality point, when the electron and hole densities are equal, this compensation leads to nonsaturating magnetoresistance due to the absence of a Hall effect  \cite{ziman}.

A more profound understanding of the mechanism behind the substantial compensating magnetoresistance in 2D semimetals has been achieved through rigorous theoretical analysis \cite{alekseev}. Their findings highlight that, particularly in narrow channels, electron-hole recombination at the boundary significantly contributes to the observed large magnetoresistance. Furthermore, under the influence of a sufficiently strong magnetic field, the boundary region can surpass the bulk contribution and result in linear magnetoresistance. Intriguingly, this effect is not confined to semimetals alone but extends to 2D topological insulators as well \cite{gusev}. It is worth noting, however, that in macroscopic samples, where a substantial number of electron-hole recombinations is anticipated in the boundary region, it is unlikely to exert a significant influence on magnetoresistivity. In such instances, the anticipated behavior of magnetoresistivity is expected to follow a parabolic trend.

Large and giant magnetoresistance can arise from sample inhomogeneity, which frequently occurs due to impurities within the material. Graphene provides a more recent example where this type of magnetoresistance has been studied \cite{furer}. Notably, significant  magnetoresistance has been observed near the charge neutrality point, where spatial inhomogeneity is caused by the two-dimensional space being divided into regions of electron and hole puddles \cite{furer, xin, levchenko}.

The Hall resistance in a two-subband system shows strong modifications with magnetic field. At low fields, the response is dominated by high-mobility carriers, while at high fields the Hall resistance converges to the classical value determined by the total carrier density \cite{ziman, zaremba, kvon}. This crossover behavior reflects the competing contributions from different subbands.

Previous studies have primarily focused on two-component systems with parabolic dispersion. In contrast, investigating a hybrid system combining carriers with both parabolic and linear (Dirac-like) spectra offers significant interest. Such a system is expected to exhibit distinct transport properties, potentially enhancing Hall and magnetoresistance effects. Detailed magnetotransport analysis enables the extraction of characteristic transport parameters for both carrier types and reveals their interplay, providing insights into novel transport phenomena.

In this study, we explore the magnetoresistance and Hall effect in a 6.3 nm gapless HgTe-based quantum well, characterized by a unique coexistence of linear (Dirac-like) and parabolic energy bands at low energies. This hybrid band system exhibits significant positive magnetoresistance and a 10 times enhanced Hall effect, driven by distinct transport properties, including differing mobilities and effective masses. Using a classical low-field magnetoresistance model that accounts for intervalley scattering, we extract key transport parameters, revealing the interplay between carriers with contrasting spectral characteristics.

\section{Electron spectrum in a gapless $HgTe$-based quantum well}
HgTe-based quantum wells have attracted considerable attention owing to their ability to generate unconventional two-dimensional (2D) systems, such as 2D topological insulators and semimetals \cite{konig, hasan, kvon3, gusev6}. The behavior of the spectrum is predominantly dictated by the thickness of the well, resulting in diverse phases characterized by insulating gaps, gapless regions, and inverted subbands \cite{gerchikov,kane, bernevig, bernevig2}. Quantum wells with a thickness of $d_{c}=6.3-6.4 nm$ are anticipated to exhibit gapless semiconductor properties with a single-valley Dirac cone near zero energy.

To explore the transport characteristics of charge carriers in a gapless HgTe quantum well, we first present the energy spectrum across a broad energy range for both the conduction and valence bands in Figure 1c. This HgTe well accommodates Dirac fermions with a linear electron and hole spectrum, denoted as $\varepsilon_e=\pm v|p|$, where the Fermi velocity is $v=7\times10^{7} cm/s = c/430$ (c represents the speed of light), and p is the momentum. Additionally, a lateral maximum of the valence band is observed below the charge neutrality point: $p_0=m_h v_F=\sqrt{2m_h(\mu-\Delta)}$, where $v_F$ is the Fermi velocity of the heavy holes, $m_h \approx 0.15 m_{0}$ is the effective mass of the holes, $\mu$ is the electrochemical potential, and $\Delta \approx 15 meV$ is indirect gap for the heavy holes (see figure 1b). Realistic samples exhibit in-plane fluctuations of the quantum well width about its average value $d\approx d_{c}$, stemming from the unavoidable variations during the quantum well growth. These fluctuations lead to in-plane variations of the band gap \cite{gusev8}, subsequently causing variations in the charge neutrality point (CNP) and minimum of the heavy energy holes positions. According to a topological network model described in ref. \cite{gusev8}, the estimated density of states (DOS) broadening is on the order of $\sim 1-4 meV$.

Figure 1d depicts the typical resistance as a function of the gate voltage (sample A). A resistance peak is evident, with the maximum corresponding to the case where the chemical potential crosses the charge neutrality point at zero energy. Below the CNP, the chemical potential $\mu$ extends into the side-located heavy hole bands. The chemical potential pinning in the tail of the heavy-hole density of states results in a pronounced asymmetry in the $\rho (V_{g})$ dependence, in contrast to graphene. This pinning may also contribute to the broadening of the density of states due to disorder. For a significant broadening of $\sim 6 meV$, the Fermi level does not intersect heavy holes energy minimum because it becomes pinned in the disorder-induced heavy hole DOS tail.

\section{Sample description and methods}
We conducted resistance measurements on quantum wells composed of $Cd_{0.65}Hg_{0.35}Te/HgTe/Cd_{0.65}Hg_{0.35}Te$ with (013) surface orientations and a well width ranging from 6.3 to 6.4 nm. The samples were epitaxially grown using Molecular Beam Epitaxy at temperatures approximately between $160^{\circ}$ and $200^{\circ}$ on GaAs substrates with (013) surface orientation (fig 1a) \cite{mikhailov}. The utilization of substrates inclined towards the 100 orientation was aimed at enhancing material quality. Details of the layer sequence scheme and sample preparation have been previously published \cite{gusev7}. Two samples (A and B) from different substrates were studied.

Experimental devices took the form of Hall bars, featuring eight voltage probes distributed across three separate segments with a width $W$ of $50 \mu$m and lengths $L$ of 100, 250, and 100 $\mu$m between the probes. A dielectric layer (200 nm of $SiO_{2}$) was deposited on the sample surface, followed by a TiAu gate electrode. Ohmic contacts to the quantum well were established by annealing indium on the device contact pads (fig 1b). The sheet density variation with gate voltage was $0.95\times10^{11}$ cm$^{-2}$V$^{-1}$. Due to dielectric breakdown, the limiting gate voltage was $\pm 8 V$. Resistance measurements $R(T)$ were carried out in the temperature range of 4.2 to 50 K using a conventional 4-probe setup with a 1-27 Hz AC current of 1-10 nA passing through the sample.
The current I flows between contacts 1 and 6, and the voltage V was measured between probes 4 and 5, $R=R_{xx} = R_{1,6}^{4,5} =V_{4,5}/I_{1,6}$ (Fig. 1b). The Hall effect was measured in configuration  $R_{xy}=R_{1,6}^{4,8} =V_{4,8}/I_{1,6}$. Resistivities are defined as $\rho_{xx}=\frac{W}{L}R_{xx}$, $\rho_{xy}=R_{xy}$.

\section{Experimental results}
\begin{figure}
\includegraphics[width=9cm]{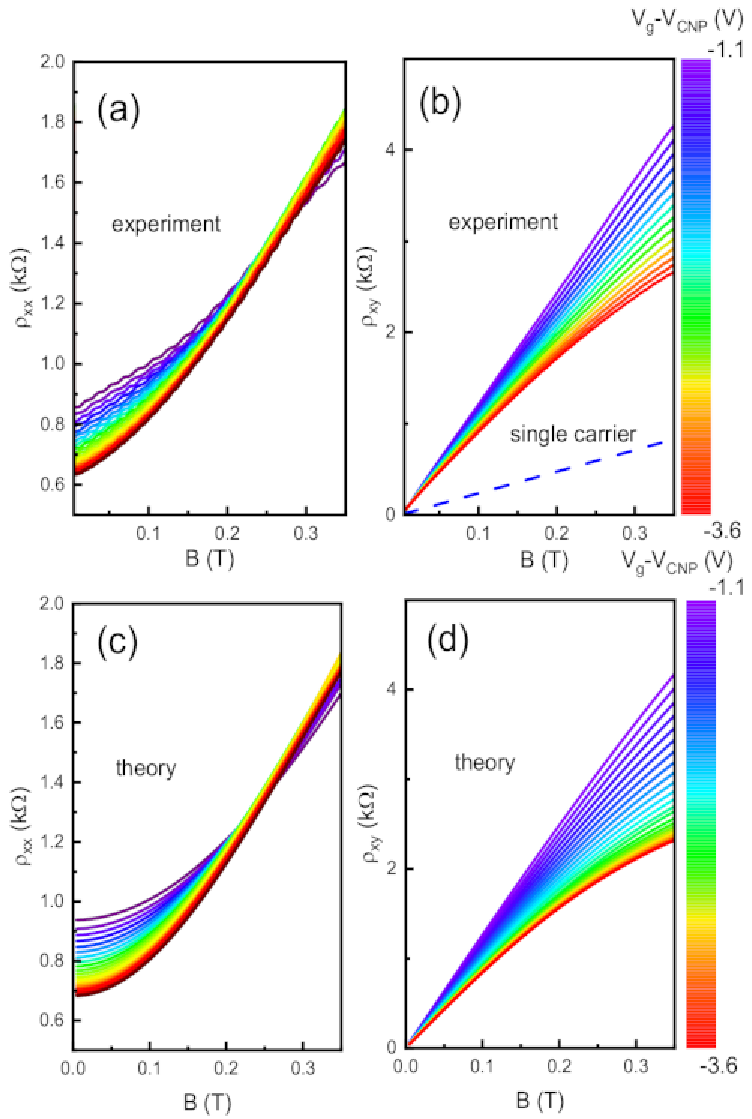}
\caption{ Gate voltage dependence of magnetoresistivity (a) and Hall effect (b) at T = 9 K in the hole transport regime (sample A). The gate voltage was varied in 0.1 V steps. The dashed line represents the Hall resistance calculated using the single-carrier model.
Panels (c) and (d) show the calculated B dependence of the magnetoresistivity and Hall effect at different gate voltages, respectively, based on Equations (1)–(4). The fitting parameters used in the calculations are presented in Fig. 4.}
\label{Magnetoresistance}
\end{figure}
\begin{figure}
\includegraphics[width=8.5cm]{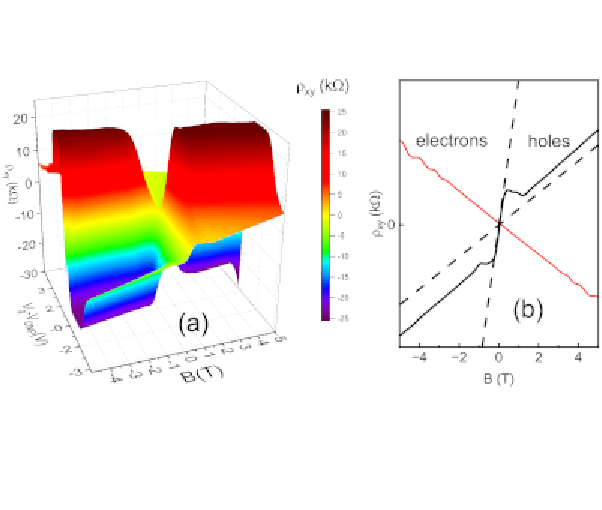}
\caption{ Hall resistivity $\rho_{xy}$ as a function of gate voltage and magnetic field at $T = 4.2\,\text{K}$.
    (b) Hall resistivity for electron and hole densities $N \approx P \approx 5.7\times10^{11}\,\text{cm}^{-2}$ (sample A). Dashed  lines show the low-field ($B \approx 0$) and high-field ($B \rightarrow \infty$) approximations for the Hall resistivity.  The ratio $\rho_{xy}(B \approx 0)/\rho_{xy}(B \rightarrow \infty) \approx 10$ is observed for holes. }
\label{Hall}
\end{figure}
\label{Magnetoresistance}

\begin{figure}
\includegraphics[width=8.5cm]{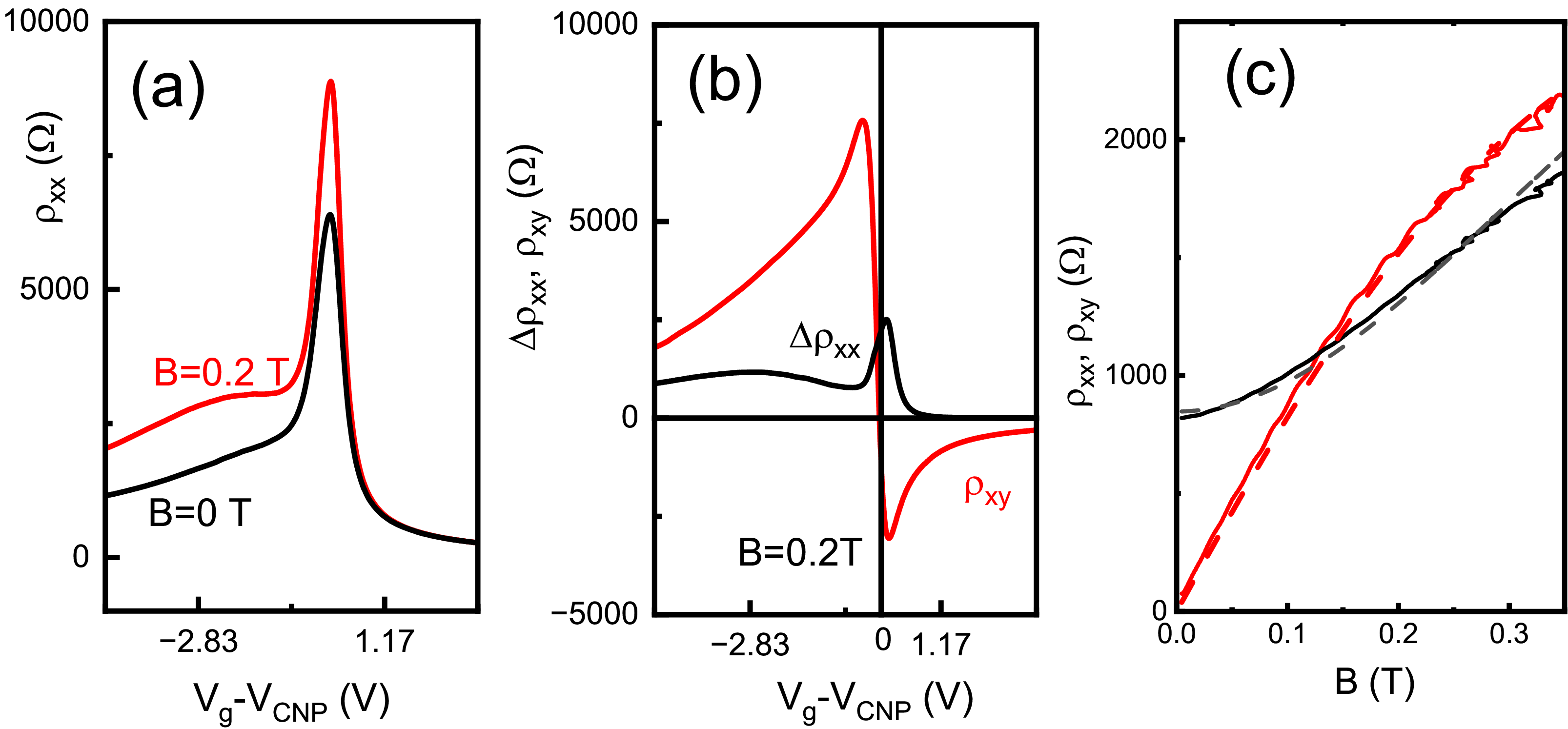}
\caption{ (a) Longitudinal resistivity $\rho_{xx}$ as a function of gate voltage at $B = 0$~T and $B = 0.2$~T for $T = 4.2$~K (sample B). The magnetoresistivity $\Delta \rho_{xx}$ and Hall resistivity $\rho_{xy}$ are also shown as functions of gate voltage at $B = 0.2$~T.
(c) Magnetoresistivity and Hall resistivity for Dirac and heavy hole densities of $P_h \approx 3.5 \times 10^{11}\,\text{cm}^{-2}$ and $Pd \approx 0.2 \times 10^{11}\,\text{cm}^{-2}$ (sample B). Dashed lines represent the theoretical fits using Equations (1)–(4) with fitting parameters: $\mu_h = 1~\text{m}^2/\text{Vs}$, $\mu_d = 18.8~\text{m}^2/\text{Vs}$, and $r = 0.2$.}
\label{Hall}
\end{figure}

Figure~1d illustrates the gate-voltage dependence of the resistance in a typical gapless HgTe well device (sample A). Over the complete gate-voltage range, the chemical potential spans a wider interval, from approximately $+150~\text{meV}$ in the electron Dirac region down to $\mu \approx 0$ at CNP, to $\mu \approx -15 meV$  at the top of the hybrid hole spectrum, and further to $-17~\text{meV}$ in the heavy hole states (Fig.~1d).

Here, we focus on the gate regime characterized by dominant hole-hole scattering. In this region, the gate voltage varies within the interval $-5.5 < V_{g} - V_{\text{CNP}} < - 0.2$; however, the chemical potential $\mu$ only varies within a narrow energy interval of $-17~\text{meV} < \mu < -15~\text{meV}$. This limited variation is a result of the high density of the heavy hole states, which is more than an order of magnitude larger than that of the Dirac holes \cite{buttner, kozlov, gusev6, kristopenko, shuvaev}.

Consequently, our system exhibits the coexistence of degenerate Dirac holes and non-degenerate heavy holes at elevated temperatures, where $ \mu - \Delta < kT < \mu $, defining a partially degenerate regime (PD) \cite{gusev5}. In this regime, we observe that the resistivity follows a temperature-dependent relationship described by a power law of $\sim T^{\alpha}$ ($\alpha \approx 3)$, attributed to scattering of the heavy holes following Boltzmann statistics by the remaining degenerate Dirac holes. Deviations from the $T^{3}$ law have been observed at low temperatures, attributed to the heavy holes becoming degenerate at lower temperatures, which causes the resistance to follow a $T^{2}$ dependence \cite{gusev5}.

In the paper  \cite{gusev5}, we present evidence that at elevated temperatures, the resistivity driven by interaction surpasses the resistivity induced by impurity scattering by a factor of 5–6. Given this observation, it is reasonable to anticipate that the magnetoresistance at low temperatures arises from the interplay between the Dirac and heavy holes subbands and interband and intraband scattering mechanisms. Thus, the previously observed temperature dependence of resistivity provides strong support for the hybrid band structure of our system.

Figure 2 illustrates the evolution of resistivity  (a) and Hall effect  (b) with magnetic field and gate voltage at a fixed temperature of T=9 K. Notably, a robust positive magnetoresistance exceeding $100 \%$ is evident on the hole side of the gate voltage dependence, particularly towards to  the charge neutrality point (CNP). Surprisingly, the measured Hall resistivity significantly exceeds conventional single-carrier predictions. The experimental Hall coefficient $R_H = \rho_{xy}/B$ is nearly an order of magnitude larger than the classical value $1/Pe$, where $P$ is total hole density,  indicating substantial deviation from standard transport behavior (fig 2b).

To investigate this anomalous behavior systematically, we performed detailed Hall resistivity measurements spanning a wide range of magnetic fields (up to 5~T) and gate voltages ($-5$ to $+5$~V), encompassing both electron- and hole-dominated transport regimes. Figure~3 displays these results as a 3D surface plot, revealing the full evolution of the Hall response across the parameter space. The most striking feature is the nearly symmetric high-field behavior, where the Hall resistivity develops a well-defined plateau at $R_{xy} = \rho_{xy}= 25.8\,\mathrm{k\Omega}$—matching the quantum resistance $h/e^2$ within experimental uncertainty ($\pm 0.5\%$)—indicating the formation of a robust quantum Hall state. We find excellent agreement between the carrier density measured via the high magnetic field Hall effect and the density calculated from the applied gate voltage. Figure 3(b) displays two representative traces of $\rho_{xy}(B)$ measured at matched electron and hole densities ($N \approx P \approx 5.7 \times 10^{11}$ cm$^{-2}$). While the electron $\rho_{xy}(B)$ follows the expected linear dependence with a Hall coefficient matching the gate-voltage-derived density, the hole transport reveals anomalous behavior: an unusually steep low-field slope transitions abruptly to a conventional slope at higher fields. Notably, the high-field hole Hall coefficient matches the magnitude (but opposite sign) of the electron value.

For reference, thin lines indicate the low-field ($B \approx 0$) and high-field ($B \rightarrow \infty$) approximations, with their ratio $\rho_{xy}(0)/\rho_{xy}(\infty) \approx 10$ highlighting the dramatic field-dependent response. This significant deviation from single-carrier transport physics strongly suggests the need for a two-subband model to properly capture the system's behavior.

Figure~4 presents the resistivity measurements for Sample~B, providing insights into its magnetotransport properties. Figure~4(a) depicts the gate voltage dependence of the longitudinal resistivity ($\rho_{xx}$) at zero magnetic field ($B = 0$) and at a magnetic field of $B = 0.2\,\text{T}$. Figure~4(b) illustrates the Hall resistivity ($\rho_{xy}$) and the magnetoresistance ($\Delta \rho_{xx} = \rho_{xx}(B) - \rho_{xx}(0)$) at $B = 0.2\,\text{T}$. A pronounced positive magnetoresistance ($\Delta \rho_{xx} > 0$) is observed in the gate voltage range where hole transport dominates, while in the electron-dominated transport regime, the magnetoresistance is minimal or negligible, consistent with observations for Sample~A. Figure~4(c) shows the magnetic field dependence of the magnetoresistance and Hall effect at a fixed hole density, revealing trends closely aligned with those observed for Sample~A, confirming the robustness of these transport characteristics across samples.

\section{Theory and Discussions}

Current theoretical models of magnetoresistance and Hall effects in two-subband systems employ the Boltzmann equation, accounting for intersubband scattering \cite{zaremba, mamani}. At low temperatures, elastic scattering from remote ionized impurities dominates, inducing both intrasubband (within a single subband) and intersubband (between subbands) transitions. The relative rates of these processes are critical. In double quantum well systems, intersubband scattering peaks at resonance, where carriers occupy both wells equally. This occurs when tunneling time is shorter than scattering time, with the total impurity-scattering rate determined by the low-mobility well, leading to an observed resistivity peak \cite{palevski, fletcher, mamani, pagnossin}. In our hybrid band system, we analyze intervalley scattering, distinguishing "Dirac-heavy hole intervalley scattering" (between Dirac and heavy-hole carriers) from "heavy-hole intervalley scattering" (between heavy-hole valleys with non-zero k).
\begin{figure}
\includegraphics[width=8.5cm]{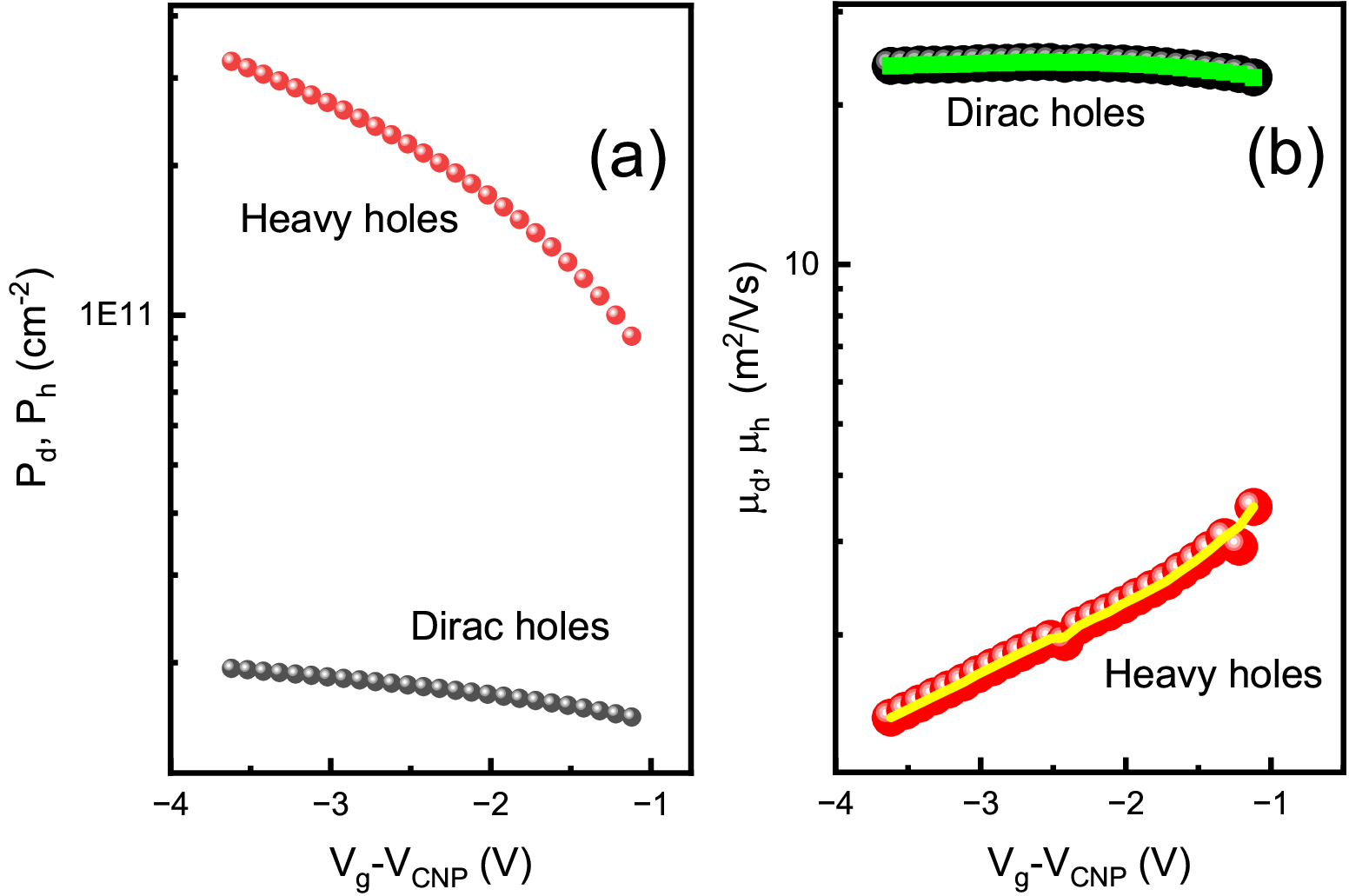}
\caption{ (a) Dirac and heavy hole carrier densities as a function of gate voltage.
(b) Dirac and heavy hole mobilities as a function of gate voltage.
Solid lines represent the theoretical fits obtained from equations (10)–(16).}
\label{Hall}
\end{figure}
 We characterize the resistivities using distinct notations for each carrier type: massless Dirac holes (denoted by subscript $p$) obey the linear dispersion $\varepsilon_p = vp$, while massive holes (denoted by $k$) follow the parabolic dispersion $\varepsilon_k = k^2/2m_h + \Delta$, with all energies referenced from the Dirac point. The corresponding cyclotron frequencies are $\omega_p = eBv^2/\mu$ for Dirac holes and $\omega_k = eB/m_h$ for massive holes, reflecting their fundamentally different responses to magnetic fields. The total zero-field conductivity $\sigma_{xx}(B=0)$ of the hybrid system combines contributions from both carrier types, expressed as $\sigma_{xx}(0) = \sigma_h + \sigma_d$. Here, $\sigma_h = (e^2 P_h/m_h)\tau_k$ represents the conductivity of massive holes with effective mass $m_h$ and scattering time $\tau_k$, while $\sigma_d = (e^2 P_d \tau_p v)/p_0$ describes Dirac holes with Fermi velocity $v$ and scattering time $\tau_p$. The characteristic momentum scale $p_0$ is determined by the Dirac hole density $P_d$ through $p_0^2 = 4\pi \hbar^2 P_d/g_d$, where $g_d = 2$ accounts for the spin degeneracy. To establish a consistent description of transport properties, we introduce effective mobilities and masses for both carrier types. For heavy holes, we define the mobility $\mu_h = e\tau_k/m_h$, yielding the conductivity $\sigma_h = P_h e\mu_h$ where $P_h$ is the heavy hole density.  The Dirac holes are characterized by mobility $\mu_d = e\tau_p v/p_0$, giving the conductivity $\sigma_d = P_d e\mu_d$ with $P_d$ as the Dirac carrier density. This unified notation reveals the analogous roles of $\mu_h$ and $\mu_d$ as the fundamental mobility parameters for their respective carriers.
 The magnetotransport properties of a hybrid system containing both heavy holes and Dirac holes can be expressed through the following relations. The longitudinal resistivity $\rho_{xx}(B)$ and Hall resistivity $\rho_{xy}(B)$, incorporating intersubband scattering effects, take the form:
\begin{widetext}
\begin{equation}
\rho_{xx}(B) = \frac{1}{e(P_h\mu_h + P_d\mu_d)} \left[1 + \frac{r P_h P_d \mu_h \mu_d (\mu_h - \mu_d)^2 B^2}{(P_h\mu_h + P_d\mu_d)^2 + r^2 P^2 \mu_h^2 \mu_d^2B^2}\right],
\label{eq:rho_xx}
\end{equation}
\end{widetext}
\begin{equation}
\rho_{xy}(B) = -\frac{\langle \mu^2 \rangle + (r\mu_h\mu_d B)^2}{\langle \mu \rangle^2 + (r\mu_h\mu_d B)^2} \cdot \frac{B}{Pe},
\label{eq:rho_xy}
\end{equation}

where we have introduced the mobility averages:

\begin{equation}
\langle \mu \rangle \equiv \frac{P_d\mu_d + P_h\mu_h}{P_d + P_h},
\label{eq:mu_avg}
\end{equation}

\begin{equation}
\langle \mu^2 \rangle \equiv \frac{P_d\mu_d^2 + P_h\mu_h^2}{P_d + P_h}.
\label{eq:mu2_avg}
\end{equation}

The dimensionless parameter $r$ quantifies the strength of intervalley scattering. In the limiting case where $r = 1$, the system reduces to the conventional two-band model of independent carriers. This behavior has been experimentally observed in GaAs-based double quantum wells, where $r \approx 1$ under off-resonance conditions as the wells become effectively decoupled, and $r<<1$ under on-resonance  conditions \cite{fletcher, mamani}. For our system, we anticipate $r \ll 1$ due to the significant disparity between the density of states of Dirac holes and heavy holes.

The magnetoresistivity exhibits a characteristic field dependence, showing parabolic behavior at low magnetic fields:

\begin{equation}
\frac{\Delta\rho_{xx}(B)}{\rho(0)} = \frac{aB^2}{1 + bB^2},
\end{equation}

where the coefficient $a = r(\mu_h - \mu_d)^2 \sigma_h \sigma_d/\sigma_0^2$ depends on the mobility contrast between heavy ($\mu_h$) and Dirac ($\mu_d$) holes, and $b = r^2 P^2 \mu_h^2 \mu_d^2/(P_h\mu_h + P_d\mu_d)^2$ reflects the intersubband scattering strength.

The Hall resistivity shows distinct limiting behaviors: the zero-field Hall coefficient $R_H(0) = -\langle \mu^2 \rangle/(\langle \mu \rangle^2 Pe)$ crosses over to the high-field limit $R_H(\infty) = -1/(Pe)$, where $P$ is the total carrier density and $\langle \mu \rangle$ represents the mobility average. Indeed $R_H(0) >> R_H(\infty)$ is expected, when  $\langle \mu^2 \rangle >> \langle \mu \rangle^2$.

Next, we fit Equations (1) and (4) to describe the zero-field resistivity, magnetoresistivity, and Hall effect. The fitting procedure involves three parameters: $\mu_d$, $\mu_h$, and $r$. The hole densities are calculated from the total carrier density, which is determined based on the gate voltage dependence and the density of states in the valence band, as described in Ref.~\cite{gusev5}, and shown in Fig.~5(a). The results of the fitting are presented in Figs.~2(c,d). We observe excellent agreement with the theoretical model proposed in Ref.~\cite{zaremba} over a broad range of magnetic fields and carrier densities. The comparison with the theoretical model is also shown in Figure 4c for sample B, demonstrating the reproducibility across different samples.

Figure~5(b) shows the gate voltage dependence of the fitting parameters for sample A. A striking contrast in the mobilities of the two carrier types is evident, indicating a dominant contribution from Dirac holes to the transport properties, including the enhanced Hall resistivity. Figure~6(a) shows the gate voltage dependence of  parameter $r$, which characterizes interband scattering. The notable feature is the small value of the parameter $r \approx 0.10$--$0.15$, suggesting a significant role of intervalley scattering. This behavior is attributed to the large difference in the density of states between the linear and parabolic branches of the energy spectrum.

\section{Intervalley and intravalley scatterign estimations}
\begin{figure}
\includegraphics[width=8.5cm]{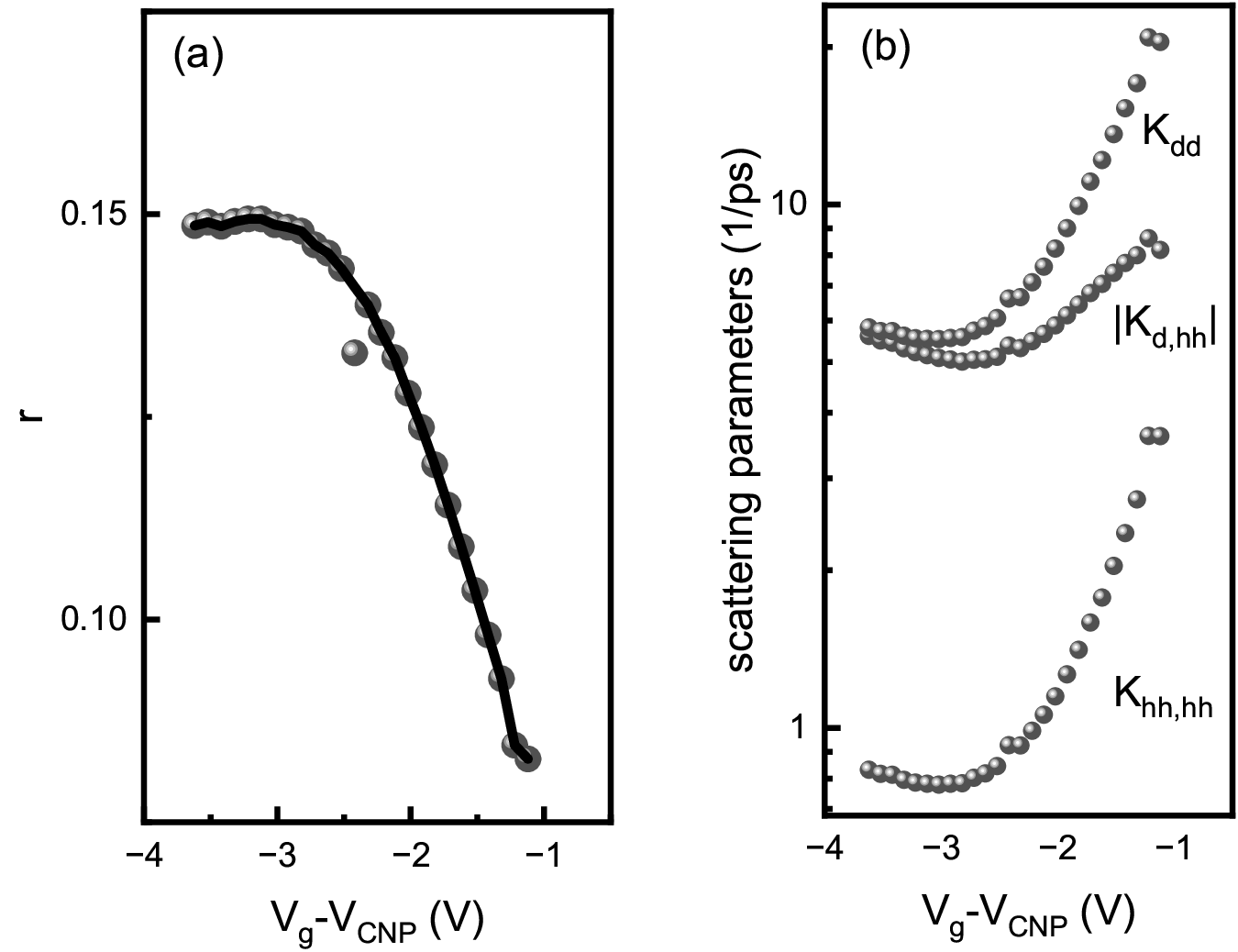}
\caption{ (a) The parameter r characterizes interband scattering. The solid line represents the theoretical fits obtained using equations (10)–(16).
(b) Transport parameters were obtained by fitting the transport scattering time and the intervalley scattering parameters using equations (10)–(16). Several constraints were applied, as explained in the text.}
\label{Hall}
\end{figure}
To analyze the small value of the parameter r obtained from magnetotransport experiments, we estimate the relevant scattering times. For simplicity, we consider a short-range potential to describe defects like vacancies and adatoms. This potential can be modeled as a delta function:  $V(\mathbf{r}) = V_0 \sum_i \delta(\mathbf{r} - \mathbf{R}_i)$. The intervalley scattering rate $\Gamma_{inter}=\tau_{\text{inter}}^{-1}$ is given by:
\begin{equation}
\frac{1}{\tau_{\text{inter}}(\epsilon)} = \frac{2\pi}{\hbar} \sum_{\mathbf{k}'} \left| \langle \mathbf{k}' | V | \mathbf{k} \rangle \right|^2 \delta(\epsilon_{\mathbf{k}'} - \epsilon_{\mathbf{k}})
\end{equation}

 where  V is scattering potential due to impurities or defects, $\mathbf{k}$ is initial state (valley $\mathbf{L}$), $\mathbf{k}'$ is final state (valley $\mathbf{L}'$).  The matrix element is given by:
$|\langle \mathbf{k}' | V | \mathbf{k} \rangle|^2 = n_{\text{imp}} |V_0|^2 |F_{\mathbf{L},\mathbf{L}'}|^2$,
where $n_{\text{imp}}$ defect or impurity density, $F_{\mathbf{L},\mathbf{L}'}$ is overlap integral between valley $\mathbf{L}$ and valley $\mathbf{L}'$. For parabolic heavy hole bands  the intervalley scattering rate is
$\tau_{\text{inter}}^{-1} = \frac{n_{\text{imp}} m_h |V_0|^2}{4\pi \hbar^3} |F_{\mathbf{L},\mathbf{L}'}|^2$, where the overlap factor $|F_{\mathbf{L},\mathbf{L}'}|^2 \sim 0.1\text{--}1$, depending on orbital symmetry. Typically for parabolic systems: $\frac{\tau_{\text{inter}}}{\tau_{\text{intra}}} = \frac{1}{|F_{\mathbf{L},\mathbf{L}'}|^2}$.
Based on this, we account for the intervalley scattering rates between Dirac and heavy holes.
The scattering process from the Dirac to the heavy hole valley ($d \rightarrow hh$) is dominated by short-range defects (e.g., vacancies, impurities), as it requires a large momentum transfer.

Conversely, the reverse channel, corresponding to the scattering of heavy holes into Dirac states ($hh \rightarrow d$), is less probable. This suppression is due to phase space constraints, specifically the limited availability of final Dirac states at low energy.

The intervalley scattering rate from Dirac to heavy holes, $\Gamma_{d,hh}$, can be estimated as:
\begin{equation}
    \Gamma_{d,hh} \approx \frac{2\pi}{\hbar} n_{\text{imp}} |V_0|^2 |F_{d,hh}|^2 \left( \frac{m_{hh}}{\pi \hbar^2} \right)
\end{equation}
where $n_{\text{imp}}$ is the impurity concentration, $V_0$ is the scattering potential strength, $F_{d,hh}$ is the wave-function overlap integral, and $m_{hh}$ is the heavy hole effective mass.

For the reverse process, the scattering rate is proportional to the energy-dependent density of states (DOS) for Dirac holes:
\begin{equation}
    \Gamma_{hh,d}(\epsilon) \propto N_d(\epsilon) = \frac{\epsilon}{\pi \hbar^2 v^2}
\end{equation}
This leads to a notable asymmetry in the scattering. The rate for $d \rightarrow hh$ processes is expected to increase with the Dirac-hole Fermi energy, $\epsilon_{F,d}$, due to the greater number of available initial states. In contrast, $hh \rightarrow d$ processes are suppressed at low $\epsilon_{F,d}$ because there are few final Dirac states to scatter into.

The scattering matrix K describes transitions between Dirac and heavy-hole states.
\begin{equation}
K_{ij} = \sum_{k \in \{d,hh\}} \Gamma_{ik}^0 \delta_{ij} - \Gamma_{ij}^1
\end{equation}

Scattering matrix elements from the matrix are:

\begin{equation}
\mathbf{K} =
\begin{pmatrix}
\Gamma_{dd}^0 + \Gamma_{d,hh}^0 - \Gamma_{dd}^1 & -\Gamma_{d,hh}^1 \\
-\Gamma_{hh,d}^1 & \Gamma_{hh,hh}^0 + \Gamma_{hh,d}^0 - \Gamma_{hh,hh}^1
\end{pmatrix}
\end{equation}

where $\Gamma_{ij}^0$ is total scattering rate from band \(i\) to \(j\), such as $\Gamma_{ij}^1$ - momentum-weighted scattering rate (accounts for backscattering).

The relaxation rates can be found from the following system of the equations\cite{zaremba}:

\begin{equation}
\tau_d^t = \frac{K_{hh,hh} - K_{d,hh} \left( \frac{\epsilon_{F,hh}}{\epsilon_{F,d}} \right)^{1/2} }{K_{dd} K_{hh,hh} - K_{d,hh} K_{hh,d}}
\end{equation}
\begin{equation}
\tau_{hh}^t = \frac{K_{dd} - K_{hh,d} \left( \frac{\epsilon_{F,d}}{\epsilon_{F,hh}} \right)^{1/2} }{K_{dd} K_{hh,hh} - K_{d,hh} K_{hh,d}}
\end{equation}

with corresponding Fermi energies for Dirac and heavy holes $\epsilon_{F,d}, \epsilon_{F,hh}$.

The eigenvalues \(\bar{\tau}_1^{-1}=\lambda_1\) and \(\bar{\tau}_2^{-1}=\lambda_2\) of the scattering matrix \(\mathbf{K}\) are given by:
\begin{widetext}
\begin{equation}
\bar{\tau}_{1,2}^{-1} = \frac{1}{2} \left( K_{dd} + K_{hh,hh} \right) \pm \frac{1}{2} \sqrt{\left( K_{dd} - K_{hh,hh} \right)^2 + 4 K_{d,hh} K_{hh,d}}
\end{equation}
\end{widetext}

Based on these equations we can write intervalley scattering parameter
\begin{equation}
r= \frac{\bar{\tau}_1 \bar{\tau}_2}{\tau_d^t \tau_{hh}^t}.
\end{equation}

With a set of simplifying assumptions for the intervalley relaxation rates, we calculate the parameters required to evaluate the intervalley mixing strength \( r \).
Considering Dirac and massive holes we obtain a system of linear equations where the mobilities $\mu_d$ and $\mu_h$ are the unknowns \cite{zaremba}.
\begin{equation}
\mu_d = \frac{e \left( k_h K_{d,hh} - k_d (\lambda_2 - K_{dd}) \right)^2}{2 \pi P_d m_d \lambda_1 \left( (K_{dd} - K_{hh,hh})^2 + 4 K_{d,hh} K_{hh,d} \right)}
\end{equation}
\begin{equation}
\mu_h = \frac{e \left( k_d (\lambda_1 - K_{dd}) - k_h K_{d,hh} \right)^2}{2 \pi P_h m_h \lambda_2 \left( (K_{dd} - K_{hh,hh})^2 + 4 K_{dd} K_{hh,hh}\right)}
\end{equation}

where the Dirac and heavy-hole momenta are given by
\[
k_d = \sqrt{2\pi P_d}, \qquad k_h = \sqrt{\pi P_h},
\]
and the effective mass of Dirac carriers is
\[
m_d = \frac{\epsilon_{F,d}}{v^2}.
\]

These are the effective mobilities of each carrier type, renormalized by the intraband $(K_{dd}, K_{hh})$ and interband scattering $(K_{d,hh}, K_{hh,d})$.
As an additional constraint, we impose strong backscattering for the intervalley parameters, assuming
\[
\frac{K_{hh,d}}{K_{d,hh}}=\frac{\Gamma_{hh,d}}{\Gamma_{d,hh}} = \frac{\epsilon_{F,d}}{m_{h} v^2} \approx 0.04 ,
\]

These three parameters $K_{dd}, K_{hh}$ and $K_{d,hh}$ allow us to fit three dependencies: the density dependence of the mixing strength \( r \) (Fig.~6a) and of $\mu_d$ and $\mu_h$ (Fig.~5b). Figure~6(a) presents the gate-voltage dependence of the fitting parameter \( r \), while Fig.~6(b) shows the corresponding behavior of the transport scattering parameters. The fitting curves, \( r(V_g) \) and $\mu_d$ and $\mu_h$, exhibit good agreement with the experimental data.

Considering the simplified nature of our model, we conclude that this level of agreement is satisfactory. A more sophisticated theoretical framework, incorporating all microscopic scattering mechanisms and a detailed transport theory, would be required for a fully quantitative description, but such an analysis lies beyond the scope of the present experimental study.

We believe, however, that our results provide a useful foundation and may stimulate further theoretical developments in the study of transport phenomena in hybrid band systems.

\section{Conclusion}
The study on the 6.3 nm gapless HgTe quantum well (QW) concludes that this two-dimensional hybrid band system, featuring coexisting linear (Dirac-like) and parabolic hole energy bands, exhibits a remarkable tenfold enhancement of the Hall resistance and large positive magnetoresistance, primarily driven by the dominant transport contribution of high-mobility Dirac holes. Using a classical two-subband model that incorporates intervalley scattering, the magnetoresistivity and Hall effect are accurately described across a broad range of magnetic fields and carrier densities.
The HgTe QW study establishes a robust framework for understanding mixed carrier magnetotransport in hybrid band systems, with the two-subband model providing a simple yet powerful tool to capture complex dynamics. Extending this framework to other hydbrid band system, such as three-dimensional topological insulators and Weyl semimetals promises to reveal how linear and parabolic dispersions, intersubband scattering, and topological effects shape transport in diverse systems. By adapting the many-subband model and addressing system-specific challenges, these studies could uncover novel transport phenomena, enhance our understanding of hybrid band systems, and pave the way for advanced quantum  devices.

\section{ACKNOWLEDGMENTS}
This work is supported by FAPESP (São Paulo Research  Foundation) Grants No. 2019/16736-2 and No. 2021/12470- 8, CNPq (National Council for Scientific and Technological  Development). The HgTe quantum wells growth and preliminary transport measurements are supported by Russian Science Foundation (Grant No. 23-72-30003 ). Discussions and correspondence with V. M. Kovalev are gratefully acknowledged.
\section{Data Availability}
The data used to generate the figures presented in this study have been deposited in the Zenodo repository and are publicly available under reference \cite{zenodo}


\begin{thebibliography}{999}
\bibitem{niu}
Rui Niu and W. K. Zhu, Materials and possible mechanisms of extremely large magnetoresistance: a
review, Journal of Physics: Condensed Matter J. Phys.: Condens. Matter {\bf 34} 113001 (2022).

\bibitem{ziman}
J. M. Ziman, Principles of the Theory of Solids (Cambridge University Press, London, 1964), Chap. 7.
\bibitem{zaremba}
E. Zaremba, Transverse magnetoresistance in quantum wells with multiple subband occupancy, Phys. Rev. B {\bf 45}, 14143 (1992).
\bibitem{fletcher}
 R. Fletcher, M. Tsaousidou, T. Smith, P. T. Coleridge, Z. R. Wasilewski, and Y. Feng,
 Two-band electron transport in a double quantum well, Phys. Rev. B {\bf 71}, 155310 (2005).
\bibitem{mamani}
N. C. Mamani, G. M. Gusev, E. C. F. da Silva, O. E. Raichev, A. A. Quivy, and A. K. Bakarov, Classical and quantum magnetoresistance in a two-subband electron system, Phys. Rev. B {\bf 80}, 085304 (2009).
\bibitem{raichev}
O. E. Raichev, Magnetic oscillations of resistivity and absorption of radiation in quantum wells with two populated subbands, Phys. Rev. B {\bf 78}, 125304 (2008).
\bibitem{kvon}
Z. D. Kvon, b, E. B. Olshanetsky, D. A. Kozlov, N. N. Mikhailov, and S. A. Dvoretskii, Two-Dimensional Electron–Hole System in a HgTe-Based Quantum Well, JETP Letters, 2008, {\bf 87},  502 (2008),
\bibitem{ali}
Mazhar N. Ali, Jun Xiong, Steven Flynn, Jing Tao, Quinn D. Gibson , Leslie M. Schoop, Tian Liang, Neel Haldolaarachchige, Max Hirschberger, N. P. Ong and R. J. Cava, Large, non-saturating magnetoresistance in $WTe_{2}$, Nature {\bf 514}, 205 (2014).
\bibitem{liang}
Tian Liang, Quinn Gibson, Mazhar N. Ali, Minhao Liu, R. J. Cava and N. P. Ong, Ultrahigh mobility and giant magnetoresistance in the Dirac semimetal $Cd_3As_2$, Nature Materials {\bf 14}, 280 (2015).
\bibitem{shekhar}
Chandra Shekhar, Ajaya K. Nayak, Yan Sun, Marcus Schmidt, Michael Nicklas, Inge Leermakers,
Uli Zeitler, Yurii Skourski, Jochen Wosnitza, Zhongkai Liu, Yulin Chen, Walter Schnelle,
Horst Borrmann, Yuri Grin, Claudia Felser and Binghai Yan, Extremely large magnetoresistance and ultrahigh mobility in the topological Weyl semimetal candidate NbP, Nature Physics {\bf 11}, 645 (2015).
\bibitem{wang}
Yaojia Wang, Lizheng Wang, Xiaowei Liu, Heng Wu, Pengfei Wang, Dayu Yan, Bin Cheng, Youguo Shi, Kenji Watanabe, Takashi Taniguchi, Shi-Jun Liang, and Feng Miao, Direct Evidence for Charge Compensation-Induced Large Magnetoresistance in Thin $WTe_{2}$, Nano Lett., {\bf 19}, 3969 (2019).
\bibitem{alekseev}
P. S. Alekseev, A. P. Dmitriev, I. V. Gornyi,  V. Yu. Kachorovskii, B. N. Narozhny, M. Schütt, and M. Titov, Magnetoresistance in Two-Component Systems, Physical Review Lett. {\bf 114}, 156601 (2015).
\bibitem{gusev}
G. M. Gusev, E. B. Olshanetsky, Z. D. Kvon, N. N.
Mikhailov, and S. A. Dvoretsky, Linear magnetoresistance in HgTe quantum wells, Phys. Rev. B {\bf 87}, 081311(R) (2013).
\bibitem{furer}
Jinglei Ping, Indra Yudhistira, Navneeth Ramakrishnan, Sungjae Cho,
Shaffique Adam, and Michael S. Fuhrer, Disorder-Induced Magnetoresistance in a Two-Dimensional Electron System, Phys. Rev. Lett. {\bf 113}, 047206 (2014).
\bibitem{xin}
Na Xin, James Lourembam, Piranavan Kumaravadivel, A. E. Kazantsev, Zefei Wu, Ciaran Mullan, Julien
Barrier, Alexandra A. Geim, I. V. Grigorieva, A. Mishchenko, A. Principi, V. I. Fal’ko, L. A. Pono-
marenko, A. K. Geim, and Alexey I. Berdyugin, Giant magnetoresistance of Dirac plasma in high-mobility graphene, Nature {\bf 616}, 270 (2023).
\bibitem{levchenko}
Alex Levchenko, Songci Li, and A. V. Andreev, Giant magnetoresistance in weakly disordered non-Galilean invariant conductors, Phys. Rev. B {\bf 109}, 075401 (2024).

\bibitem{buttner}
 B. Buttner, C. X. Liu, G. Tkachov, E. G. Novik, C. Brne,
H. Buhmann, E. M. Hankiewicz, P. Recher, B. Trauzettel,
S. C. Zhang, and L. W. Molenkamp, Single valley Dirac fermions in zero-gap HgTe quantum wells, Nat. Phys. 7, 418 (2011).
\bibitem{kozlov}
D. A. Kozlov, Z. D. Kvon, N. N. Mikhailov, and S. A. Dvoretskii, Weak localization of Dirac fermions in HgTe quantum wells, JETP Lett. {\bf 96}, 730 (2012).
\bibitem{gusev6}
G. M. Gusev, D. A. Kozlov,  A. D. Levin,  Z. D. Kvon,  N. N. Mikhailov,  and S. A. Dvoretsky, Robust helical edge transport at $\nu=0$ quantum Hall state,
Phys. Rev. B {\bf 96}, 045304 (2017).
\bibitem{kristopenko}
S. S. Krishtopenko, W. Desrat, K. E. Spirin, C. Consejo, S. Ruffenach, F. Gonzalez-Posada, B. Jouault, W. Knap, K. V. Maremyanin, V. I. Gavrilenko, G. Boissier, J. Torres, M. Zaknoune, E. Tournié, and F. Teppe,
Massless Dirac fermions in III-V semiconductor quantum wells, Phys. Rev. B {\bf 99} 121405(R) (2019)

\bibitem{shuvaev}
Alexey Shuvaev, Vlad Dziom , Jan Gospodaric , Elena G. Novik, Alena A. Dobretsova,
Nikolay N. Mikhailov, Ze Don Kvon and Andrei Pimenov, Band Structure Near the Dirac Point in HgTe Quantum Wells with Critical Thickness,  Nanomaterials {\bf 12}, 2492 (2022).
\bibitem{konig}
M. Konig, S. Wiedmann, C. Brune, A. Roth, H. Buhmann,
L. W. Molenkamp, X.-L. Qi, and S.-C. Zhang,  Quantum spin Hall insulator state in HgTe quantum wells. Science {\bf 318}, 766,  1148047 (2007)
\bibitem{hasan}
M. Z. Hasan, M. Z.  and C. L.  Kane, Colloquium: Topological insulators. Rev. Mod. Phys. {\bf 82}, 3045  (2010).
\bibitem{kvon3}
Z. D. Kvon, D. A. Kozlov, E. B. Olshanetsky, G.M.Gusev, N. N. Mikhailov, S. A. Dvoretsky, Topological insulators based on HgTe. Phys.-Usp. {\bf 63}, 629
(2020).
\bibitem{gusev7}
G.M. Gusev, Z.D. Kvon, E. B. Olshanetsky, N. N. Mikhailov, Mesoscopic transport in two-dimensional topological insulators. Solid State Commun., {\bf 302}, 113701 (2019).
\bibitem{gerchikov}
L. G. Gerchikov and A. V. Subashiev, Interface States in Subband Structure
of Semiconductor Quantum Wells, phys. stat. sol. (b) {\bf 160}, 443 (1990).
\bibitem{kane}
 C. L. Kane and E. J. Mele, $Z_{2}$ Topological Order and the Quantum Spin Hall Effect, Phys. Rev. Lett. {\bf 95}, 146802 (2005).
\bibitem{bernevig}
B. A. Bernevig and S. C. Zhang, Quantum Spin Hall Effect, Phys. Rev. Lett. {\bf 96}, 106802 (2006);
\bibitem{bernevig2}
B. A. Bernevig, T. L. Hughes, and S. C. Zhang, Quantum Spin Hall Effect and Topological Phase Transition in HgTe Quantum Wells, Science {\bf 314}, 1757 (2006).
\bibitem{gusev8}
G. M. Gusev, Z. D. Kvon, D. A. Kozlov, E. B. Olshanetsky, M. V. Entin and N. N. Mikhailov, Transport through the network of topological channels in HgTe based quantum well, 2D Mater. {\bf 9}  015021 (2022).
\bibitem{mikhailov}
N.N. Mikhailov, R.N. Smirnov, S.A. Dvoretsky, Yu.G. Sidorov, V.A. Shvets, E.V. Spesivtsev and S.V. Rykhlitski, Growth of $Hg_{1-x}Cd_{x}Te$ nanostructures by molecular beam epitaxy with ellipsometric control, International Journal of Nanotechnology, {\bf 3}, 120 (2006).

\bibitem{gusev5}
G. M. Gusev,  A. D. Levin,  E. B. Olshanetsky,  Z. D. Kvon,  V. M. Kovalev, M. V. Entin,  and N. N. Mikhailov, Interaction-dominated transport in two-dimensional conductors:
From degenerate to partially degenerate regime, Phys. Rev. B {\bf 109},  035302  (2024).

\bibitem{palevski}
Alexander Palevski, Fabio Beltram, Federico Capasso, Loren Pfeiffer, and Kenneth W. West,
Resistance Resonance in Coupled Potential Wells, Phys. Rev. Lett. {\bf 65}, 1929  (1990).
\bibitem{pagnossin}
I. R. Pagnossin, A. K. Meikap, T. E. Lamas, G. M. Gusev, and J. C. Portal, Anomalous dephasing scattering rate of two-dimensional electrons in double quantum well structures, Phys. Rev. B {\bf 78}, 115311 (2008).
\bibitem{zenodo}
G. M. Gusev. (2025).Data set : Longitudinal and Hall Resistance of HgTe quantum well as a Function of Temperature, Magnetic field and Gate Voltage. Zenodo. https://doi.org/10.5281/zenodo.15863358.
\end{thebibliography}
\end{document}